\documentclass[a4paper]{article}

\usepackage{INTERSPEECH2019}
\usepackage{CJK}
\usepackage{amssymb,bm}
\usepackage{textcomp}
\usepackage{amsmath}
\usepackage{epstopdf}
\usepackage{longtable}
 \usepackage{rotating}
 \usepackage{subfigure}
 \usepackage{graphicx}
\usepackage{multicol}
\usepackage{booktabs}
\usepackage{algorithm,algpseudocode,amsfonts}
\usepackage{algorithmicx}
\usepackage{multirow}
\usepackage{float}
\usepackage{bm}

\def\vec#1{\mathchoice%
	{\mbox{\boldmath $\displaystyle\bf#1$}}
	{\mbox{\boldmath $\textstyle\bf#1$}}
	{\mbox{\boldmath $\scriptstyle\bf#1$}}
	{\mbox{\boldmath $\scriptscriptstyle\bf#1$}}}
\def\vc#1{\protect\vec #1}

\usepackage{pifont}
\usepackage{subfigure}
\usepackage{mathrsfs}    
\usepackage{verbatim}
\usepackage{siunitx}
\usepackage{cite}
\usepackage{siunitx}
\usepackage{dblfloatfix}
\usepackage{array}
\usepackage{pifont,xcolor}
\usepackage{hyperref}
\usepackage[export]{adjustbox}
\usepackage{multirow,bigstrut}
\def\vec#1{\ensuremath{\bm{{#1}}}}

\ninept

\title{The DKU Replay Detection System for the ASVspoof 2019 Challenge: \\
On Data Augmentation, Feature Representation, Classification, and Fusion}

\name{Weicheng Cai$^{1, 2}$, Haiwei Wu$^{1, 2}$, Danwei Cai$^1$, and Ming Li$^1$\thanks{This research was funded in part by the National Natural Science Foundation of China (61773413), Natural Science Foundation of Guangzhou City (201707010363), Six talent peaks project in Jiangsu Province (JY-074), Science and Technology Program of Guangzhou City (201903010040), and Huawei. We thank Weixiang Hu, Yu Lu, Zexin Liu and Lei Miao from Huawei Digital Technologies Co., Ltd, China who provided insight and expertise that greatly assisted this research.}}
\address{
  $^1$Data Science Research Center, Duke Kunshan University, Kunshan, China\\
  $^2$School of Electronics and Information Technology, Sun Yat-sen University, Guangzhou, China}
\email{ming.li369@dukekunshan.edu.cn}

\begin{document}

  \maketitle
\begin{abstract}
This paper describes our DKU replay detection system for the ASVspoof 2019 challenge. The goal is to develop spoofing countermeasure for automatic speaker recognition in physical access scenario.  We leverage the countermeasure system pipeline from four aspects, including the data augmentation, feature representation, classification, and fusion. First, we introduce an utterance-level deep learning framework for anti-spoofing. It receives the variable-length feature sequence and outputs the utterance-level scores directly. Based on the framework, we try out various kinds of input feature representations extracted from either the magnitude spectrum or phase spectrum. Besides, we also perform the data augmentation strategy by applying the speed perturbation on the raw waveform. Our best single system employs a residual neural network trained by the speed-perturbed group delay gram. It achieves EER of 1.04\% on the development set, as well as  EER of 1.08\% on the evaluation set. Finally, using the simple average score from several single systems can further improve the performance. EER of 0.24\% on the development set and 0.66\% on the evaluation set is obtained for our primary system.
\end{abstract}

\section{Introduction}

Automatic speaker verification (ASV) refers to automatically accept or reject a claimed identity by analyzing speech utterances, and nowadays it is widely used in real-world biometric authentication applications~\cite{Reynolds1995Robust, Kinnunen2010An, Hansen2015Speaker}. Recently, a growing number of studies have confirmed the severe vulnerability of state-of-the-art ASV systems under a diverse range of intentional fraudulent attacks on various databases~\cite{Evans2013Spoofing, Wu2015Spoofing, Ergunay2015On}. The initiative of the series ASVspoof challenge aims to promote the development of spoofing countermeasure studies~\cite{ZhizhengASVspoof}. The task in previous ASVspoof 2015 challenge was to discriminate genuine human speech from speech produced using text-to-speech and voice conversion attacks~\cite{Wu2015ASVspoof}. The ASVspoof 2017 challenge is to assess audio replay spoof attack detection ``in the wild", and the spoofing recordings are created from real uncontrolled replay setup~\cite{Kinnunen2017RedDots, Kinnunen2017, Delgado2018}.

The ASVspoof 2019 challenge extends the previous challenge in several directions~\cite{asvspoof2019}. In this paper, we only focus on the replay detection sub-task in the physical access scenario. A controlled setup in the form of replay attacks simulated using a range of real replay devices and carefully controlled acoustic conditions are adopted. The goal is to develop countermeasure systems that can distinguish between bona fide and the spoofed speech by replay. 

 Recently, the Gaussian Mixture Model~(GMM) classifier trained with constant Q cepstral coefficient~(CQCC) feature has been the benchmark for various anti-spoofing tasks~\cite{Hanil2015Classifiers, Todisco2016A}. The CQCC feature is a perceptually-inspired time-frequency analysis extracted from a constant-Q transform (CQT)~\cite{Brown1991Calculation, TODISCO2017516}. The GMM is an unsupervised generative model and is widely used to depict the probability distribution of audio features. It assumes that all frames of features are independent and identically distributed, and features are grouped to estimate the parameters of GMM in the training stage. Given the GMM classifier, the final utterance-level scores are derived by averaging the frame-level log-likelihood altogether. 
 
 Although the CQCC-GMM official baseline is an effective spoofing countermeasure, there remains much potential to improve. Previous work in~\cite{7815339, Lavrentyeva2017, Tom:2018dn} have investigated the efficiency of deep learning approach compared to the GMM classifier. Various kinds of feature representation other than CQCC, such as short-time Fourier transform~(STFT) gram~\cite{Lavrentyeva2017}, group delay gram~(GD~gram)~\cite{Tom:2018dn} are also explored and show superior performance. Besides, Cai~\textit{et al.} have also demonstrated that suitable data augmentation~(DA) can also significantly improve the replay detection system performance~\cite{Cai2017}. 
 
 Taking the DKU system for ASVspoof 2019 challenge as the pivot, we aim to explore the countermeasure system in physical access scenario from the following four aspects, including the data augmentation, feature representation, classification, and fusion. First, we introduce an utterance-level deep learning framework for anti-spoofing. Considering the variable-length nature of speech,  the deep neural network~(DNN) receives a feature sequence of arbitration duration and outputs the utterance-level posteriors directly. Based on the framework, we try out various kinds of input feature representations from either magnitude spectrum or phase spectrum. They include the CQCC, linear frequency cepstral coefficients~(LFCC), inverted Mel-frequency cepstral coefficients~(IMFCC), STFT gram, and GD gram. Besides, we also perform a simple yet effective DA strategy by applying the speed perturbation on the raw waveform to increase the quantity of training data.

\begin{figure*}[tb]
        \centering    
        \includegraphics[width=0.96\textwidth]{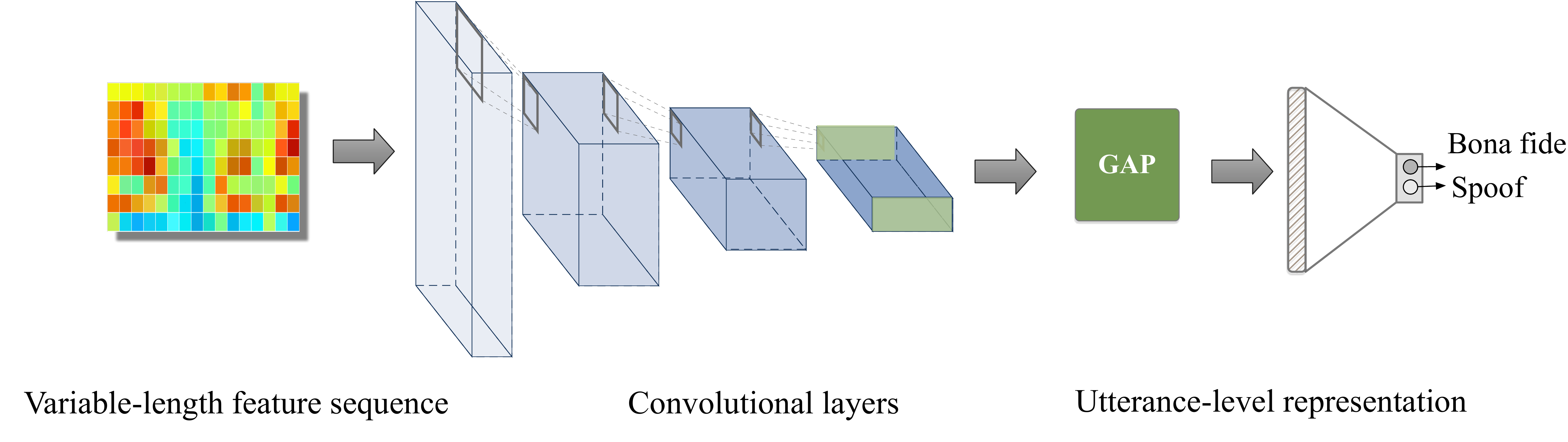}
        \caption{Utterance-level DNN framework for anti-spoofing.  It accepts input data sequence with variable length, and  produces an utterance-level result directly from the output of the DNN.}\label{fig:spoof_framework}
 \end{figure*}  
 
\section{Methods}
\subsection{Utterance-level DNN framework}
\label{sec:framework}

All of our systems are built upon a unified utterance-level DNN framework. We first use the approach for the task of speaker and language recognition in previous works~\cite{Cai_2018_Odyssey, caie2e_iccasp18, Cai_2018_Interspeech}. As demonstrated in Fig.~\ref{fig:spoof_framework}, the DNN framework accepts variable-length feature sequence and produces an utterance-level result from the output unit directly.

The network structure here is somewhat similar to that one in~\cite{Tom:2018dn}. However, there exist two main differences: (\lowercase\expandafter{\romannumeral1}) the input feature sequence is first either truncated or padded to a fixed length along the time axis, and then further resized to a $512 \times 256$ ``image" before feeding into the DNN in~\cite{Tom:2018dn}. In contrast, we employ a variable-length training strategy, and in the testing stage, the raw full-length feature sequences are fed into our DNN directly. (\lowercase\expandafter{\romannumeral2}) the training procedure in~\cite{Tom:2018dn} contains the complicated two-stage flow. Both of the two-stage networks rely on a pre-trained model from the large-scale Imagenet~\cite{deng2009imagenet} dataset. In contrast, we train a single DNN from scratch only using the ASVspoof 2019 training set.

We first transform the raw waveform into a frame-level feature sequence based on hand-crafted filters. The output sequence of feature extraction is a matrix of size $D \times L$, where $D$ is the feature dimension along the frequency axis, and $L$ denotes the frame length along the time axis.

As depicted in Fig.~\ref{fig:spoof_framework}, we use a deep convolutional neural network~(CNN) to further transform the raw feature into a high-level abstract representation. For a given feature sequence of size $D \times L$, typically, the CNN learned descriptions are a three-dimensional tensor block of shape $C\times H\times W$, where $C$ denotes the number of channels, $H$ and $W$ denotes the height and width of the feature maps. Generally, $H$ and $W$ are much smaller than the original $D$ and $L$, since we have many downsample operations within the CNN structure. 

The CNN acts as a local pattern extractor, and the learned representation by CNN is still with temporal order. The remaining question is: how to aggregate the whole sequence together over time? Concerning about that, we adopt a global average pooling~(GAP) layer on the top of CNN~\cite{LinNiN}. Given CNN learned feature maps $\vc {F} \in \mathbb{R}^{C\times H\times W}$,  the GAP layer accumulates mean statistics along with the time--frequency axis, and the corresponding output is defined as:
\begin{equation}
v_i =\frac{1}{H\times W}\times  \sum_{j=1}^{j=H}\sum_{k=1}^{k=W} \vc {F}_{i,j,k}
\end{equation}
 
Therefore, we get fixed-dimensional utterance-level representation $\vc {V}={\left[v_1, v_2, \cdots, v_C\right]}$ from the output of GAP. 

We further process the utterance-level representation through a fully-connected feed-forward network and build an output layer on top. The two units in the output layer are represented as bona fide and spoof categories. We can optimize the whole countermeasure system in an end-to-end manner with a cross-entropy loss. The final utterance-level score can be directly fetched from the DNN output. 

\subsection{Feature representation}
In this section, we investigate different input feature representations upon the introduced DNN framework in Section~\ref{sec:framework}.

\subsubsection{CQCC}
The CQCC feature is obtained by perceptually-aware CQT coupled with traditional cepstral analysis. It is reported to be sensitive to the general form of spoofing attack, and yields superior performance among various kinds of features~\cite{TODISCO2017516}. More details of CQCC can be found in \cite{Todisco2016A}.

\subsubsection{LFCC}
The official baseline systems adopt the CQCC feature as well as the LFCC feature.

LFCC is a kind of cepstral features based on triangle filterbank similar to the widely-used Mel-frequency cepstral coefficients (MFCC). It is extracted the same way as MFCC, but the filters are in the same triangular shape rather than on mel scale. Therefore, LFCC might have better resolution in the higher frequency region~\cite{Sahid2015}.

\subsubsection{IMFCC}

IMCC is also a kind of filter bank based cepstral features.  The difference with MFCC is that IMFCC uses filters that are linearly placed in ``inverted-mel" scale, which lays more stress on the high-frequency region~\cite{Sahid2015}.

\subsubsection{STFT gram}

Let $x(n)$ be a given speech sequence and   $X_n(\omega)$ its STFT after applying a window $w(n)$ on the speech signal  $x(n)$. $X(\omega)$  can be expressed as  
\begin{equation}
X_n(\omega) = \vert X_n(\omega)\vert e^{j\theta_n(\omega)}.   
\end{equation}
where $\vert X_n(\omega)\vert$ corresponds to the short-time magnitude spectrum and $\theta_n(\omega)$ corresponds to the phase spectrum.

The square of the magnitude spectrum is called the STFT power spectrum. We adopt the logarithm of the power spectrum as the STFT gram.

\subsubsection{Group delay gram}

Most spectral features are derived from the STFT magnitude spectrum, while the short-time phase spectrum is not used, considering that the fact that the human ear is phase ``deaf". However, the phase spectrum has been used for various speech processing tasks including synthetic speech detection and audio replay detection recently~\cite{1198718, Hegde:ju, SARATXAGA201630, Tom:2018dn}.   Paliwal \textit{et al.} have shown that deviations in the phase that are not linear are important for perception~\cite{PALIWAL2005153},  and deviations in phase can be represented using the group delay function.
\begin{equation}
\tau(\omega) = -\frac{d(\theta(\omega))}{d\omega}       
\end{equation}
where the phase spectrum $\theta(\omega)$ of a signal is defined as a continuous function of $\omega$. The group delay function can also be computed from the signal as in~\cite{Hegde:ju} using 

 \begin{equation}
 \begin{aligned}
\tau_x(\omega) & =  -  \Im   \left[ \frac{d(\log({X(\omega)}))}{d\omega} \right]\\
& =  \frac{X_{\Re}(\omega)Y_{\Re}(\omega) + Y_{\Im}(\omega)X_{\Im}(\omega)}{{\vert X(\omega)\vert}^2} 
\end{aligned}      
 \end{equation}
 where $\Re$ and $\Im$ refer to the real and imaginary parts of the Fourier transform. $X(\omega)$ and $Y(\omega)$ are the Fourier transforms of $x(n)$ and $nx(n)$, respectively.

\subsubsection{Joint gram}

Regarding that the speech signal is completely characterized by both the short-time magnitude and phase spectrum, here we combine that STFT gram and GD gram together to form a new feature representation. Typically, we can consider the input feature sequence as a $1\times D \times T$ ``image", where 1 denotes the image channel, $D$ denotes the image height, $T$ denotes the image width. Thus the joint gram is formed as a two-channel ``image". For example, given separated STFT gram and JD gram of size $512\times T$ from the same utterance,  a joint gram of size $2\times512\times T$ is formed as the DNN input.
    
\subsection{Data augmentation}

DA, which is a common strategy adopted to increase the quantity of training data, has been shown to be effective for neural network training to make robust predictions. Many methods such as the x-vector system rely on external datasets/resource to perform data augmentation. However, external speech data is strictly forbidden for the ASVspoof 2019 challenge. Considering that, we adopt the DA strategy by applying a simple speed perturbation. Specifically, we apply standard 3-way speed perturbation using factors of 0.9, 1.0 and 1.1~\cite{Tom2015}. After that, we have training data of 3 times the original for both bona fide and spoof category.

\subsection{Model selection and fusion}

Ideally, we can use the development dataset to select the best DNN checkpoint and then train the ensemble parameter on the development set for fusion. However, the evaluation plan indicates that we should submit both development and evaluation set scores. If we do so, the performance on the development set will be overfitting and lead to a rather low error rate.

Regarding this fact, we treat the development set the same as the evaluation set. This means that we only use the training set for both DNN training and system fusion. Since we have no separated validation set, the converged model after the last optimization step is selected for evaluation. As for the system fusion, we only adopt a simple score-level average operation to get the final ensemble score. 

\newcommand{\blocka}[2]{\multirow{2}{*}{\(\left[\begin{array}{c}\text{3$\times$3, #1}\\[-.1em] \text{3$\times$3, #1} \end{array}\right]\)$\times$#2}
}
\newcommand{\blockb}[3]{\multirow{2}{*}{\(\left[\begin{array}{c}\text{3$\times$3, #2}\\[-.1em] \text{3$\times$3, #1}\end{array}\right]\)$\times$#3}
}
\renewcommand\arraystretch{1.1}
\setlength{\tabcolsep}{3pt}
\begin{table}[t]
    \caption{ Detailed network structure and number of parameters for each module. It downsamples at the last three blocks (Res2, Res3, and Res4) using stride=2 for their first convolutional layer. The total number of parameters is about 1.33M.}
        \centering
    \begin{adjustbox}{max width=0.99    \columnwidth}
        \begin{tabular}{|c|c|c|c|c|c|c|c|c|c|c|}
            \hline
            Layer& Input size & Output size & Structure&\#Params \\
            \hline
            Conv1 & $ 1\!\times\!512\!\times\!L$ & $\!16\!\times\!512\!\times\!L$& $3\!\times\!3$, stride 1&144    \\
            \hline
            \multirow{2}{*}{Res1} & \multirow{2}{*}{$16\!\times\!512\!\times\!L$} &  \multirow{2}{*}{$16\!\times\!512\!\times\!L$} &  \blockb{16}{16}{3}    &    \multirow{2}{*}{ $4K\times 3$ } \\
            &&&&\\
            \hline
            \multirow{2}{*}{Res2} & \multirow{2}{*}{$16\!\times\!512\!\times\!L$} &  \multirow{2}{*}{$32\!\times\!256\!\times\!\frac{L}{2}$} &    \blockb{32}{32}{4}&\multirow{2}{*}{$18K\times 4$}  \\
            & & & &\\\hline
            \multirow{2}{*}{Res3} & \multirow{2}{*}{ $32\!\times\!256\!\times\!\frac{L}{2}$} &  \multirow{2}{*}{$64\!\times\!128\!\times\!\frac{L}{4}$} &   \blockb{64}{64}{6}  &    \multirow{2}{*}{$72K \times 6$} \\         
            & & & &\\
            \hline
            \multirow{2}{*}{Res4} & \multirow{2}{*}{$64\!\times\!128\!\times\!\frac{L}{4}$} &  \multirow{2}{*}{$128\!\times\!64\times\!\frac{L}{8}$} &   \blockb{128}{128}{3}    &    \multirow{2}{*}{$    288K\times 3$} \\         
            & & & &\\\hline
            
            GAP& $128\!\times\!64\!\times\!\frac{L}{8}$&128& Pooling  &0\\
   
            \hline
            FC & $128$&$ 32$ &Fully-connected & 4K\\
            \hline
            Output & $32 $&2 &Fully-connected &   66    \\
            \hline
        \end{tabular}
    \end{adjustbox}
    \label{tab:resnetconfigt}
\end{table}

\section{Experiments}

\subsection{Data protocol and evaluation metrics}

We strictly respect the official protocols defined in the evaluation plan and submit both development and evaluation set scores. 
We have \num{54000} training utterances altogether, including \num{5400} bona fide audio and \num{48600} spoofed audio from various replay conditions.

The primary metric is the minimum normalized tandem decision cost function (min-tDCF) defined by the organizer~\cite{Kinnunen2018}. An alternative equal error rate~(EER) is adopted as a secondary metric.

\subsection{Feature extraction}
The LFCC feature, CQCC feature, as well as the GMM classifier,  is exactly followed the official baseline provided by the organizer.

For IMFCC feature, 20-dimensional static coefficients are augmented with their delta and double delta coefficients, making 60-dimensional IMFCC feature vectors.

For STFT gram and GD gram, the FFT bins are set to \num{1024}. The frame length is \num{25} ms, with a frameshift \num{10} ms. Finally, we get \num{512} feature vector for each frame.

\subsection{Network setup}

To get higher-level abstract representation, we design a deep CNN based on the well-known residual neural network~(ResNet)~\cite{ He2016Deep}. We adopt the ResNet-34 backbone, but our network is slightly ``thin" that the output channels of the front-end ResNet are only up to \num{128}, which leads to a small number of total parameters. The core modules and the corresponding parameters are described in Table~\ref{tab:resnetconfigt}. We use batch normalization followed by the ReLU activation functions after every convolutional layer. The convolutional layer is all with kernels of $3\times 3$. 

We adopt the common stochastic gradient descent (SGD) algorithm with momentum \num{0.9} and weight decay 1e-4.  The learning rate is set to \num{0.1}, \num{0.01}, \num{0.001} and is switched when the training loss plateaus.  In the training stage, the model is trained with a mini-batch size of \num{128}. We design a data loader to generate the variable-length training examples on the fly. For each training step,  an integer $L$ within $\left[150 \textrm{,}  350 \right]$  interval is randomly generated, and each data in the mini-batch is truncated or extended to $L$ frames. Therefore, a dynamic mini-batch of data with the shape of $128 \times D \times L$ is generated prior to each training step, where $D$ is the input feature dimension, and $L$ is a batch-wise variable number indicating the frame length. 

We choose the output from the bona fide unit as our final score. In the testing stage, the full-length feature sequence is directly fed into the network, without any truncate or padding operation. Due to the fact that utterances may have arbitrary durations, we feed the testing audio to the trained network one by one.

\begin{table}[t]
\caption{Comparison of different classifiers  on the development set }
\label{tab: classifier}
\centering
 \resizebox{0.7\columnwidth}{!}{    
\begin{tabular}[c]{l l  c c}
\toprule
 \textbf{Feature} &\textbf{Classifier}&\textbf{EER(\%)}& \textbf{min-tDCF} \\
 \toprule
 CQCC & GMM&9.87&0.1953\\
 CQCC & \textbf{ResNet}& \textbf{4.77} &\textbf{0.1127}\\
   \midrule
 LFCC&GMM& 11.96&0.2554 \\
 LFCC&\textbf{ResNet} & \textbf{2.62}&\textbf{0.0609} \\
\bottomrule
\end{tabular}}
\end{table}

\begin{table}[t]
\caption{Comparison of different types of feature  on the development set}
\label{tab: feature}
\centering
 \resizebox{0.66\columnwidth}{!}{
\begin{tabular}[c]{l c c}
\toprule
 \textbf{Feature} &\textbf{EER(\%)}& \textbf{min-tDCF} \\
 \toprule
 CQCC &4.77&0.1127\\
 LFCC & 2.62&0.0609\\
 IMFCC& 3.66&0.0890 \\
 STFT gram & 4.11&0.1070 \\
  \textbf{GD gram} & \textbf{1.81} &\textbf{0.0467} \\
\bottomrule
\end{tabular}} 
\end{table}

\begin{table}[t]
\caption{Effect of the DA strategy  on the development set}
\label{tab: da}
\centering
 \resizebox{0.69\columnwidth}{!}{
\begin{tabular}[c]{l c c c}
\toprule
 \textbf{Feature} &\textbf{DA}&\textbf{EER(\%)}& \textbf{min-tDCF} \\
 \toprule
 STFT gram& \ding{55}&4.11    &0.1070\\
 STFT gram&\textbf{\ding{51}} &\textbf{3.35} &\textbf{0.0904}\\
 \midrule
 GD gram & \ding{55}&1.81    &0.0467\\
 GD gram & \textbf{\ding{51}}&\textbf{1.03}    &\textbf{0.0265}\\
\bottomrule
\end{tabular} }
\end{table}

\begin{table}[tb]
\caption{Overall performance on both the development set and the evaluation set. EER are written before the slash, min-tDCF are behind. N/R: Not Reported.}
\label{tab: eval}
\centering
 \resizebox{0.96\columnwidth}{!}{
\begin{tabular}[c]{l c c c c c c}
\toprule
 \textbf{ID}& \textbf{System}&\textbf{DA}&\textbf{Dev}&\textbf{Eval} \\
 \toprule
 1& CQCC--GMM&\ding{55}& 9.87/0.1953&11.04/0.2454\\
 2& LFCC--GMM &\ding{55}&11.96/0.2554&13.54/0.3017\\
    \midrule
 3& LFCC--ResNet&\ding{55}& 2.62/0.0609&N/R\\
 4&   IMFCC--ResNet &\ding{55}& 3.66/0.0890&N/R\\
 5&     STFT gram--ResNet&\ding{51}& 3.35/0.0904&N/R\\
 6&     GD gram--ResNet &\ding{55}& 1.81/0.0467&1.79/0.0439\\
 7& GD gram--ResNet &\ding{51}&\textbf{1.03/0.0265}&\textbf{1.08/0.0282}\\
 8& Joint gram--ResNet &\ding{51}& 1.14/0.0209&1.23/0.0305\\   
    \midrule
\multicolumn{3}{c}{\textbf{Fusion 3+4+5+6+7+8}}& \textbf{0.24/0.0064}&\textbf{0.66/0.0168}\\
\bottomrule
\end{tabular}}
\end{table}

\subsection{Results on back-end classification}

To compare the back-end classifier, we fix the front-end feature representation as CQCC/LFCC in this section. From Table~\ref{tab: classifier},  we can see that no matter for CQCC or LFCC feature, the ResNet classifier based on our introduced utterance-level DNN framework is significantly superior to the baseline GMM classifier. Therefore, we fix the back-end classifier as the ResNet model for the following experiments.

\subsection{Results on front-end feature representation}

In this section, we fix the ResNet back-end classifiers and perform experiments on different types front-end feature representations. From Table~\ref{tab: feature}, we can see that the performance of the countermeasure system strongly depends on the choice of the feature representation. The GD gram based on the phase spectrum yields the best among the investigated five features.

\subsection{Results on data augmentation}

The results in Table~\ref{tab: da} show the effectiveness of the DA based on speed perturbation strategy. 18\% relative EER reduction is achieved on the development set for the STFT gram system, and 48\% for the GD gram system. 

\subsection{Results on fusion and generalization}

Table~\ref{tab: eval} shows the system performance on both the development and the evaluation set. Since only four systems can be submitted at most, some system performance on the evaluation set could not be reported at this stage. Our best GD gram--ResNet system achieves relative 90\% EER reduction as well as min-tDCF  on both development set and evaluation set. Finally, using the simple average score from several single systems can further improve the performance. EER of 0.24\% on the development set and 0.66\% on the evaluation set is obtained for our primary system. This reveals the complementariness of the system outputs from classifiers trained with different features.

The performance on the development set and evaluation set are very close, reveals the robustness of our proposed system.

\section{Conclusion}

In this paper, we take the DKU system for ASVspoof 2019 challenge as the pivot and explore the design of the countermeasure system against replay attack through data augmentation, feature representation, classification, and fusion. First, we utilize a ResNet back-end classifier based on the introduced utterance-level deep learning framework. Second, we investigate various kinds of front-end feature representations and find that the GD gram based on the phase spectrum is pretty effective for replay detection. Moreover, a simple data augmentation based on speed perturbation can also significantly improve the performance, especially for the GD gram--ResNet system. Finally, both the single and primary fusion systems dramatically decrease the EER and min-tDCF on not only on the development set but also on the evaluation set.

\bibliographystyle{IEEEtran}

\bibliography{spoof}

\end{document}